\documentclass[prd,aps,superscriptaddress,nofootinbib,amsmath,amssymb, 11pt]{revtex4}
\usepackage{graphicx}
\usepackage{amstext,amsfonts}
\usepackage{bm}
\usepackage{xcolor}
\usepackage{hyperref}
\usepackage{braket}
\def\slashchar#1{\setbox0=\hbox{$#1$}
   \dimen0=\wd0 \setbox1=\hbox{/} \dimen1=\wd1
   \ifdim\dimen0>\dimen1 \rlap{\hbox to \dimen0{\hfil/\hfil}} #1
   \else  \rlap{\hbox to \dimen1{\hfil$#1$\hfil}} / \fi}

\graphicspath{{./eps/}}

\newcommand{\fluc}{{fluctuations }}
\newcommand{\flucc}{{fluctuations. }}
\newcommand{\urq}{\footnotesize{URQMD}}

\begin{document}
\thispagestyle{empty}
\begin{center}
{\Large {\bf Centrality Dependence of Multiplicity Fluctuations in Ion-Ion Collisions from the Beam Energy Scan at FAIR}}\\  

	{\bf Anuj Chandra, Bushra Ali and Shakeel Ahmad\footnote{email: Shakeel.Ahmad@cern.ch\\Authors declare that there is no conflict of interest}}\\
{\it Department of Physics, Aligarh Muslim University\\  [-2mm] Aligarh INDIA}\\
[4ex]
\end{center}

\begin{center}
{\bf Abstract}\\  
\end{center}
Multiplicity distributions and event-by-event multiplicity fluctuations in AuAu collisions at energies in future heavy-ion experiment at the Facility for Anti-proton and Ion Research ({\footnotesize{FAIR}}) are investigated. Events corresponding to {\footnotesize{FAIR}} energies are simulated in the frame work of Ultra Relativistic Quantum Molecular Dynamics ({\urq}) model. It is observed that the mean and the width of multiplicity distributions monotonically increase with beam energy. The trend of variations of dispersion with mean number of participating nucleons for the centrality-bin width of 5\% are in accord with the Central Limit Theorem. The multiplicity distributions in various centrality bins as well as for full event samples are observed to obey Koba, Nielsen and Olesen ({\footnotesize{KNO}}) scaling. The trends of variations of scaled variance with beam energy are also found to support the {\footnotesize{KNO}} scaling predictions for larger collision centrality. The findings also reveal that the statistical fluctuations in 5\% centrality-bin width appear to be under control.\\ 

\noindent {\footnotesize PACS numbers: } 25.75.−q, 25.75.Gz\\ [10ex]

\noindent KEY-WORDS: Beam Energy Scan, Event-by-event fluctuations, Relativistic heavy-ion collisions.\\
\newpage
\noindent {\bf 1. Introduction}\\
\noindent Any physical quantity measured in an experiment is subject to \flucc These \fluc depend on the property of the system and are expected to provide important information about the nature of the system under study [1,2]. As regards relativistic heavy-ion (AA) collisions, the system so created is a dense and hot fireball consisting of partonic and (or) hadronic matter [1,2]. To investigate the existence of partonic matter in the early life of fireball is one of the main goals of AA collisions. Study of \fluc in AA collisions would help check the idea that \fluc of a thermal system are directly related to various susceptibilities and could be an indicator for the possible phase transitions [1,2,3]. Fluctuations in experimental observables, such as charged particle multiplicity, particle ratios, mean transverse momentum and other global observables are related to the thermodynamic properties of the system, like, entropy, specific heat, chemical potential, etc. [4,5,6,7]. Event-by-event (ebe) \fluc of these quantities are regarded as an important mean to understand the particle production dynamics which, in turn, would lead to understand the nature of phase transition and the critical \fluc at the QCD phase boundary. A non-monotonic behavior of the \fluc as a function of collision centrality and energy of the colliding beam may signal the onset of confinement and may be used to probe the critical point in the QCD phase diagram [7]. The multiplicity of charged particles produced in heavy-ion collisions is the simplest and day-one observable which provides a mean to investigate the dynamics of highly excited multi-hadron system. Studies involving multiplicity distributions (MDs) of the relativistic charged particles produced would allow finding the deviations from a simple superposition of multiple independent nucleon-nucleon ($nn$) collisions. Such studies, if carried out in limited rapidity space are envisaged to provide useful information on dynamical \fluc [8,9,10,11]. It has been stressed that moments of MDs in full and limited rapidity bins would lead to make some interesting remarks about the production mechanisms involved. Dependence of MDs and their moments on collision centrality is also expected to lead to some interesting conclusions because of the fact that in narrow centrality windows the geometrical \fluc may be treated as under control, whereas, such windows, if correspond to most central collisions, may be of additional importance because of the extreme conditions of temperature and excitation energy [7]. An attempt is, therefore, made to study the multiplicity \fluc in the narrow centrality windows in AuAu collisions for the Beam Energy Scan ({\footnotesize{BES}}) at {\footnotesize{FAIR}} energies (for $E_{lab} =$ 10, 20, 30 and 40A GeV) in the frame work of {\urq} model, using the code, urqmd-v3.4 [12, 13]. The number of events simulated at these energies are 2.3, 2.3, 2.1 and 2.2M (M = 10$^{6}$) respectively. The analysis is carried out in the pseudorapidity and p$_{T}$ intervals with $-1.0 < \eta < 1.0$ and $0.2 < p_{T} < 5.0$ GeV/c respectively. \\ 
\noindent{\bf 2. The {\footnotesize{URQMD}} Model}\\ 
\noindent Multiparticle production in relativistic collisions have been described earlier within the hydrodynamic approach [14]. At a later stage the Regge theory [15] and multiperipheral models were developed [15,16]. Although the difficulties attributed to the statistical models wereover come in these models yet the inconvenience of this approach is the large number of free parameters which are to be fixed by comparison with the experiments. Subsequently various quark-parton models motivated by QCD were introduced and as a consequence a large variety of models for hadronic and heavy-ion collisions were proposed. These models may be classified into macroscopic (statistical and thermodynamic) models [17] and microscopic (string, transport, cascade, etc.) models, like {\footnotesize{URQMD}}, {\footnotesize{VENUS}}, {\footnotesize{RQMD}}, etc. The microscopic models describe the individual hadron-hadron collisions.\\
\begin{table}[htbp]
\small
\begin{tabular}{c r r r r}\hline
\multicolumn{1}{c}{Centrality($\%$)} & \multicolumn{1}{c}{$<N_{part}>$} & \multicolumn{1}{c}{$<N_{ch}>$} & \multicolumn{1}{c}{$\sigma$} & \multicolumn{1}{c}{$\omega$}\\[0.2em]
\hline
   5  &   348.00 $\pm$ 0.0020   &  231.03 $\pm$ 0.07  &   25.07 $\pm$ 0.05  &   2.7198 $\pm$ 0.0009 \\
  10  &   289.90 $\pm$ 0.0020   &  182.37 $\pm$ 0.06  &   21.63 $\pm$ 0.05  &   2.5650 $\pm$ 0.0009 \\
  15  &   238.45 $\pm$ 0.0020   &  144.36 $\pm$ 0.05  &   19.34 $\pm$ 0.05  &   2.5898 $\pm$ 0.0010 \\
  20  &   195.27 $\pm$ 0.0020   &  113.62 $\pm$ 0.04  &   17.26 $\pm$ 0.05  &   2.6214 $\pm$ 0.0012 \\
  25  &   159.21 $\pm$ 0.0020   &   89.70 $\pm$ 0.04  &   15.19 $\pm$ 0.05  &   2.5727 $\pm$ 0.0013 \\
  30  &   127.17 $\pm$ 0.0020   &   70.10 $\pm$ 0.04  &   13.85 $\pm$ 0.05  &   2.7352 $\pm$ 0.0016 \\
  35  &   100.08 $\pm$ 0.0020   &   53.49 $\pm$ 0.03  &   12.18 $\pm$ 0.05  &   2.7734 $\pm$ 0.0019 \\
  40  &    77.97 $\pm$ 0.0010   &   39.72 $\pm$ 0.03  &   10.76 $\pm$ 0.05  &   2.9171 $\pm$ 0.0022 \\
  45  &    58.89 $\pm$ 0.0010   &   28.99 $\pm$ 0.02  &    9.22 $\pm$ 0.05  &   2.9297 $\pm$ 0.0027 \\
  50  &    44.44 $\pm$ 0.0010   &   20.62 $\pm$ 0.02  &    7.84 $\pm$ 0.05  &   2.9803 $\pm$ 0.0032 \\
  55  &    31.99 $\pm$ 0.0010   &   13.86 $\pm$ 0.02  &    6.46 $\pm$ 0.05  &   3.0073 $\pm$ 0.0039 \\
  60  &    22.27 $\pm$ 0.0010   &    9.13 $\pm$ 0.01  &    5.17 $\pm$ 0.05  &   2.9287 $\pm$ 0.0049 \\
  65  &    14.83 $\pm$ 0.0010   &    5.95 $\pm$ 0.01  &    4.14 $\pm$ 0.05  &   2.8798 $\pm$ 0.0059 \\
  70  &    10.15 $\pm$ 0.0010   &    3.67 $\pm$ 0.01  &    3.21 $\pm$ 0.05  &   2.8056 $\pm$ 0.0071 \\
  75  &     7.06 $\pm$ 0.0020   &    2.17 $\pm$ 0.01  &    2.42 $\pm$ 0.05  &   2.6991 $\pm$ 0.0089 \\
  80  &     5.53 $\pm$ 0.0050   &    1.20 $\pm$ 0.00  &    1.77 $\pm$ 0.05  &   2.6062 $\pm$ 0.0111 \\
\hline
\end{tabular}
\caption{Values of \(<N_{part}>\), \(<N_{ch}>\), dispersion ($\sigma$) and scaled variance ($\omega$) in various centrality bins at E\,$_{lab}$ = 10A GeV/c}
\end{table}
\begin{table}[htbp]
\small
\begin{tabular}{c r r r r}\hline
\multicolumn{1}{c}{Centrality($\%$)} & \multicolumn{1}{c}{$<N_{part}>$} & \multicolumn{1}{c}{$<N_{ch}>$} & \multicolumn{1}{c}{$\sigma$} & \multicolumn{1}{c}{$\omega$}\\[0.2em]
\hline 
   5  &   348.00 $\pm$ 0.0020   &   288.81 $\pm$ 0.07   &   27.91 $\pm$ 0.05  &   2.6980 $\pm$ 0.0008 \\
  10  &   289.90 $\pm$ 0.0020   &   227.70 $\pm$ 0.06   &   24.58 $\pm$ 0.05  &   2.6535 $\pm$ 0.0008 \\
  15  &   238.45 $\pm$ 0.0020   &   179.76 $\pm$ 0.06   &   22.37 $\pm$ 0.04  &   2.7839 $\pm$ 0.0010 \\
  20  &   195.27 $\pm$ 0.0020   &   141.24 $\pm$ 0.05   &   20.27 $\pm$ 0.04  &   2.9100 $\pm$ 0.0012 \\
  25  &   159.21 $\pm$ 0.0020   &   111.38 $\pm$ 0.05   &   18.10 $\pm$ 0.03  &   2.9414 $\pm$ 0.0014 \\
  30  &   127.17 $\pm$ 0.0020   &    86.96 $\pm$ 0.04   &   16.52 $\pm$ 0.03  &   3.1384 $\pm$ 0.0017 \\
  35  &   100.08 $\pm$ 0.0020   &    66.34 $\pm$ 0.04   &   14.62 $\pm$ 0.03  &   3.2211 $\pm$ 0.0020 \\
  40  &    77.97 $\pm$ 0.0010   &    49.44 $\pm$ 0.03   &   12.90 $\pm$ 0.02  &   3.3637 $\pm$ 0.0024 \\
  45  &    58.89 $\pm$ 0.0010   &    36.03 $\pm$ 0.03   &   11.05 $\pm$ 0.02  &   3.3876 $\pm$ 0.0030 \\
  50  &    44.44 $\pm$ 0.0010   &    25.75 $\pm$ 0.02   &    9.49 $\pm$ 0.02  &   3.4943 $\pm$ 0.0035 \\
  55  &    31.99 $\pm$ 0.0010   &    17.36 $\pm$ 0.02   &    7.80 $\pm$ 0.01  &   3.5043 $\pm$ 0.0042 \\
  60  &    22.27 $\pm$ 0.0010   &    11.53 $\pm$ 0.02   &    6.32 $\pm$ 0.01  &   3.4638 $\pm$ 0.0055 \\
  65  &    14.83 $\pm$ 0.0010   &     7.55 $\pm$ 0.01   &    5.11 $\pm$ 0.01  &   3.4627 $\pm$ 0.0066 \\
  70  &    10.15 $\pm$ 0.0010   &     4.71 $\pm$ 0.01   &    4.00 $\pm$ 0.01  &   3.4008 $\pm$ 0.0080 \\
  75  &     7.06 $\pm$ 0.0020   &     2.79 $\pm$ 0.01   &    3.06 $\pm$ 0.01  &   3.3583 $\pm$ 0.0100 \\
  80  &     5.53 $\pm$ 0.0050   &     2.25 $\pm$ 0.21   &    2.62 $\pm$ 0.15  &   3.0484 $\pm$ 0.3079 \\
\hline
\end{tabular}
\caption{Values of the same variables, as in Table I, but for E\,$_{lab}$ = 20A GeV/c}
\end{table}
\begin{table}[htbp]
\small
\begin{tabular}{c r r r r}\hline
\multicolumn{1}{c}{Centrality($\%$)} & \multicolumn{1}{c}{$<N_{part}>$} & \multicolumn{1}{c}{$<N_{ch}>$} & \multicolumn{1}{c}{$\sigma$} & \multicolumn{1}{c}{$\omega$}\\[0.2em]
\hline
   5  &   348.00 $\pm$ 0.0020  &    327.23 $\pm$ 0.09   &   31.63 $\pm$ 0.06  &   3.0566 $\pm$ 0.0009 \\
  10  &   289.90 $\pm$ 0.0020  &    257.60 $\pm$ 0.08   &   27.73 $\pm$ 0.05  &   2.9857 $\pm$ 0.0010 \\
  15  &   238.45 $\pm$ 0.0020  &    203.53 $\pm$ 0.07   &   25.22 $\pm$ 0.05  &   3.1245 $\pm$ 0.0011 \\
  20  &   195.27 $\pm$ 0.0020  &    159.94 $\pm$ 0.06   &   22.70 $\pm$ 0.04  &   3.2205 $\pm$ 0.0013 \\
  25  &   159.21 $\pm$ 0.0020  &    126.11 $\pm$ 0.06   &   20.26 $\pm$ 0.04  &   3.2551 $\pm$ 0.0016 \\
  30  &   127.17 $\pm$ 0.0020  &     98.62 $\pm$ 0.05   &   18.56 $\pm$ 0.03  &   3.4930 $\pm$ 0.0019 \\
  35  &   100.08 $\pm$ 0.0020  &     75.28 $\pm$ 0.04   &   16.47 $\pm$ 0.03  &   3.6032 $\pm$ 0.0023 \\
  40  &    77.97 $\pm$ 0.0010  &     56.15 $\pm$ 0.04   &   14.48 $\pm$ 0.03  &   3.7316 $\pm$ 0.0027 \\
  45  &    58.89 $\pm$ 0.0010  &     40.95 $\pm$ 0.03   &   12.40 $\pm$ 0.02  &   3.7530 $\pm$ 0.0033 \\
  50  &    44.44 $\pm$ 0.0010  &     29.36 $\pm$ 0.03   &   10.63 $\pm$ 0.02  &   3.8484 $\pm$ 0.0040 \\
  55  &    31.99 $\pm$ 0.0010  &     19.87 $\pm$ 0.02   &    8.83 $\pm$ 0.02  &   3.9211 $\pm$ 0.0048 \\
  60  &    22.27 $\pm$ 0.0010  &     13.20 $\pm$ 0.02   &    7.19 $\pm$ 0.01  &   3.9103 $\pm$ 0.0063 \\
  65  &    14.83 $\pm$ 0.0010  &      8.68 $\pm$ 0.02   &    5.81 $\pm$ 0.01  &   3.8935 $\pm$ 0.0076 \\
  70  &    10.15 $\pm$ 0.0010  &      5.44 $\pm$ 0.01   &    4.60 $\pm$ 0.01  &   3.8949 $\pm$ 0.0094 \\
  75  &     7.06 $\pm$ 0.0020  &      3.23 $\pm$ 0.01   &    3.51 $\pm$ 0.01  &   3.8103 $\pm$ 0.0116 \\
  80  &     5.53 $\pm$ 0.0050  &      3.01 $\pm$ 0.25   &    3.13 $\pm$ 0.18  &   3.2661 $\pm$ 0.2933 \\
\hline
\end{tabular}
\caption{Values of the same variables, as in Table I, but for E\,$_{lab}$ = 30A GeV/c}
\end{table}
\begin{table}[htbp]
\small
\begin{tabular}{c r r r r}\hline
\multicolumn{1}{c}{Centrality($\%$)} & \multicolumn{1}{c}{$<N_{part}>$} & \multicolumn{1}{c}{$<N_{ch}>$} & \multicolumn{1}{c}{$\sigma$} & \multicolumn{1}{c}{$\omega$}\\[0.2em]
\hline
   5  &   348.00 $\pm$ 0.0020   &   353.86 $\pm$ 0.10   &   35.23 $\pm$ 0.07   &  3.5065 $\pm$ 0.0010 \\
  10  &   289.90 $\pm$ 0.0020   &   278.50 $\pm$ 0.08   &   30.55 $\pm$ 0.06   &  3.3503 $\pm$ 0.0011 \\
  15  &   238.45 $\pm$ 0.0020   &   219.93 $\pm$ 0.07   &   27.86 $\pm$ 0.05   &  3.5289 $\pm$ 0.0013 \\ 
  20  &   195.27 $\pm$ 0.0020   &   172.92 $\pm$ 0.07   &   24.98 $\pm$ 0.05   &  3.6082 $\pm$ 0.0015 \\ 
  25  &   159.21 $\pm$ 0.0020   &   136.43 $\pm$ 0.06   &   22.09 $\pm$ 0.04   &  3.5766 $\pm$ 0.0017 \\ 
  30  &   127.17 $\pm$ 0.0020   &   106.50 $\pm$ 0.05   &   20.16 $\pm$ 0.04   &  3.8159 $\pm$ 0.0020 \\ 
  35  &   100.08 $\pm$ 0.0020   &    81.49 $\pm$ 0.05   &   17.75 $\pm$ 0.03   &  3.8673 $\pm$ 0.0024 \\ 
  40  &    77.97 $\pm$ 0.0010   &    60.80 $\pm$ 0.04   &   15.78 $\pm$ 0.03   &  4.0966 $\pm$ 0.0029 \\ 
  45  &    58.89 $\pm$ 0.0010   &    44.31 $\pm$ 0.04   &   13.50 $\pm$ 0.03   &  4.1148 $\pm$ 0.0036 \\ 
  50  &    44.44 $\pm$ 0.0010   &    31.76 $\pm$ 0.03   &   11.55 $\pm$ 0.02   &  4.1974 $\pm$ 0.0042 \\ 
  55  &    31.99 $\pm$ 0.0010   &    21.58 $\pm$ 0.02   &    9.61 $\pm$ 0.02   &  4.2835 $\pm$ 0.0051 \\ 
  60  &    22.27 $\pm$ 0.0010   &    14.31 $\pm$ 0.02   &    7.78 $\pm$ 0.01   &  4.2327 $\pm$ 0.0066 \\ 
  65  &    14.83 $\pm$ 0.0010   &     9.43 $\pm$ 0.02   &    6.28 $\pm$ 0.01   &  4.1880 $\pm$ 0.0079 \\ 
  70  &    10.15 $\pm$ 0.0010   &     5.90 $\pm$ 0.01   &    4.95 $\pm$ 0.01   &  4.1634 $\pm$ 0.0097 \\ 
  75  &     7.06 $\pm$ 0.0020   &     3.50 $\pm$ 0.01   &    3.77 $\pm$ 0.01   &  4.0707 $\pm$ 0.0120 \\ 
  80  &     5.53 $\pm$ 0.0050   &     2.04 $\pm$ 0.03   &    2.97 $\pm$ 0.02   &  4.3207 $\pm$ 0.0620 \\
\hline
\end{tabular}
\caption{Values of the same variables, as in Table I, but for E\,$_{lab}$ = 40A GeV/c}
\end{table}
\noindent {\urq} model is based on the co-variant propagation of constituent quarks and di-quarks but has been accompanied by baryonic and mesonic degrees of freedom. At low energies, $\sqrt{s_{NN}} <$ 5 GeV, the collisions are described in terms of interactions between hadrons and their excited states [17], whereas at higher energies ( $>$ 5 GeV), the quark and gluon degrees of freedom are considered and the concept of color string excitation is introduced with their subsequent fragmentation into hadrons [13]. In a transport model, AA collisions are considered as the superposition of all possible binary $nn$ collisions. Every $nn$ collision corresponding to the impact parameter, $b \leq \sqrt{\sigma_{tot}/\pi}$ is considered, where $\sigma_{tot}$ represents the total cross section. The two colliding nuclei are described by Fermi gas model [17] and hence the initial momentum of each nucleon is taken at random between zero and Thomas-Fermi momentum. The interaction term includes more than 50 baryon and 45 meson species. The model can treat the intermediate fireball both in and out of a local thermal and chemical equilibria. The {\urq} model, thus, provides an ideal framework to study heavy-ion collisions. Although, the phase transition from a hadronic to partonic phase are not explicitly included in the model, thus a clear suggestion about the location of critical point cannot be made. The study, however, might help in the interpretation of the experimental data since it will permit subtraction of simple dynamical and geometrical effects from the expected Quark Gluon Plasma ({\footnotesize{QGP}}) signals [18]. \\ 


\noindent{\bf 3. Results and discussion}\\
\noindent The {\urq} model gives the value of impact parameter, {\it b} on ebe basis which allows to determine the collision centrality and mean number of participating nucleons, $\langle N_{part}\rangle$ using the Glauber model [7,19]. Values of number of participating nucleons, mean charged particle multiplicities and dispersion of MDs ($\sigma$) for various collision centralities at the four energies are estimated and listed in Tables I - IV. The centrality selection is made from the MDs of charged particles for the minimum bias events in the considered \(\eta\) and \(p_t\) ranges. This is illustrated in FIG.1, where the multiplicity distribution of minimum bias events for E$_{lab}$ = 40A GeV is displayed. The shaded regions show 10\% centrality cross-section bins.  Variations of $\langle N_{ch}\rangle$ and $\sigma$ with $\langle N_{part}\rangle$ for the centrality bin width = 2, 5 and 10\% are presented in FIGs.2 and 3. The statistical errors associated with these parameters are too small to be noticed in the figure. It may be noted from the figure that $\langle N_{ch}\rangle$ and $\sigma$ increase smoothly with $\langle N_{part}\rangle$ or collision centrality. The lines in FIG.2 are due to the best fits to the data obtained using the  equation
\begin{eqnarray}
	<N_{ch}> = a + b<N_{part}> + c<N_{part}>^2
\end{eqnarray}
whereas, in FIG.3 the lines are due to the least square fits to the data of the form

\begin{eqnarray}
	\sigma = p + q\sqrt{<N_{part}>} 
\end{eqnarray}

\begin{table}[htbp]
\small
\vspace{0.5em}
\begin{tabular}{|l|c|r|r|r|} \hline
E$_{lab}$ & Fit Par. &Centrality 10$\%$ & Centrality 5$\%$ & Centrality 2$\%$ \\ 
\hline\hline
10A GeV & a $\times$ 10$^{-1}$ &  -17.425 $\pm$ 0.032 & -19.211 $\pm$ 0.071 & -13.718 $\pm$ 0.004  \\ \cline{2-5}
                         & b $\times$ 10$^{-2}$ &   49.595 $\pm$ 0.018 &  49.581 $\pm$ 0.020 &  43.791 $\pm$ 0.015  \\ \cline{2-5}
                         & c $\times$ 10$^{-4}$ &    4.895 $\pm$ 0.008 &   4.452 $\pm$ 0.007 &   5.350 $\pm$ 0.005  \\ \hline\hline

20A GeV & a $\times$ 10$^{-1}$ &  -15.317 $\pm$ 0.051 & -18.215 $\pm$ 0.063 & -17.031 $\pm$ 0.045  \\ \cline{2-5}
                         & b $\times$ 10$^{-2}$ &   60.103 $\pm$ 0.024 &  60.250 $\pm$ 0.021 &  53.923 $\pm$ 0.017  \\ \cline{2-5}
                         & c $\times$ 10$^{-4}$ &    6.668 $\pm$ 0.010 &   6.670 $\pm$ 0.007 &   6.900 $\pm$ 0.006  \\
\hline\hline
30A GeV & a $\times$ 10$^{-1}$ &  -16.633 $\pm$ 0.061 & -18.670 $\pm$ 0.066 & -17.852 $\pm$ 0.050  \\ \cline{2-5}
                         & b $\times$ 10$^{-2}$ &   68.297 $\pm$ 0.029 &  68.483 $\pm$ 0.024 &  60.889 $\pm$ 0.019  \\ \cline{2-5}
                         & c $\times$ 10$^{-4}$ &    7.458 $\pm$ 0.012 &   7.428 $\pm$ 0.008 &   7.872 $\pm$ 0.007  \\
\hline\hline
40A GeV & a $\times$ 10$^{-1}$ &  -18.499 $\pm$ 0.063 & -19.733 $\pm$ 0.065 & -18.947 $\pm$ 0.049  \\ \cline{2-5}
                         & b $\times$ 10$^{-2}$ &   74.059 $\pm$ 0.030 &  73.900 $\pm$ 0.026 &  65.885 $\pm$ 0.020  \\ \cline{2-5}
                         & c $\times$ 10$^{-4}$ &    7.986 $\pm$ 0.013 &   8.080 $\pm$ 0.010 &   8.468 $\pm$ 0.007  \\

\hline
\end{tabular}
\caption{Values of parameters, a, b and c, occurring in Eq.1 at different energies.}
\end{table}

\begin{table}[htbp]
\small
\vspace{0.5em}
\begin{tabular}{|l|c|r|r|r|} \hline

E$_{lab}$ & Fit Par. & Centrality 10$\%$ & Centrality 5$\%$ & Centrality 2$\%$ \\ 
\hline\hline
10A GeV & p $\times$ 10$^{-1}$ &   -21.171 $\pm$ 0.061   &  -10.181 $\pm$ 0.087  &   8.087 $\pm$ 0.055  \\ \cline{2-5}
                         & q $\times$ 10$^{-1}$ &    16.436 $\pm$ 0.017   &   13.244 $\pm$ 0.001  &  11.704 $\pm$ 0.011  \\
\hline\hline
20A GeV & p $\times$ 10$^{-1}$ &   -19.790 $\pm$ 0.082   &   -8.694 $\pm$ 0.008  &  -6.706 $\pm$ 0.067  \\ \cline{2-5}
                         & q $\times$ 10$^{-1}$ &   -19.050 $\pm$ 0.019   &   15.308 $\pm$ 0.001  &  13.634 $\pm$ 0.013  \\
\hline\hline
30A GeV & p $\times$ 10$^{-1}$ &   -21.280 $\pm$ 0.097   &   -8.795 $\pm$ 0.008  &  -6.936 $\pm$ 0.073  \\ \cline{2-5}
                         & q $\times$ 10$^{-1}$ &    21.334 $\pm$ 0.023   &   17.131 $\pm$ 0.002  &  15.295 $\pm$ 0.015  \\
\hline\hline
40A GeV & p $\times$ 10$^{-1}$ &   -22.988 $\pm$ 0.101   &  -11.275 $\pm$ 0.008  &  -8.658 $\pm$ 0.075  \\ \cline{2-5}
                         & q $\times$ 10$^{-1}$ &    23.292 $\pm$ 0.026   &   18.901 $\pm$ 0.002  &  16.814 $\pm$ 0.016  \\
\hline
\end{tabular}
\caption{Values of parameters, p and q, occurring in Eq.2 at different energies.}
\end{table}

\noindent The values of coefficients, occurring in Eqs.1 and 2 are listed in Tables V and VI respectively. As described in ref.7, the centrality dependence of the moments may be understood by the Central Limit Theorem (CLT), according to which, $\langle N_{ch}\rangle \propto N_{part}$ and $\sigma \propto \sqrt{N_{part}}$. However, in the present study the mean multiplicity is observed to grow with $\langle N_{part}\rangle$, as given by Eq.1., i.e. a slight deviation from linearity is exhibited by the data irrespective of the fact that how large or small the centrality bins are chosen. The variations of \(\sigma\) with $\langle N_{part}\rangle$, shown in FIG.3, is seen to be nicely fitted by Eq.2 for 5\% centrality bin width, while for the centrality bin widths of 2\% and 10\% the data are seen to be fitted only for centrality \(>\) 20\%, as indicated by the lines in this figure; the lines are drawn for the range of centrality for which the fits of the data have been performed. Similar deviations from CLT have also been observed in AuAu collisions at RHIC and lower energies [7]. In order to extract dynamical \fluc arising from physical processes, \fluc in mean number of participating nucleons are to be minimized. To achieve the same, centrality bins considered should be kept narrow because the \fluc in the particle multiplicities are directly related to the \fluc in the mean number of participating nucleons. The inherent \fluc may be reduced by choosing narrow centrality bins; the inherent \fluc are the \fluc which arise due to the difference in the geometry even within the selected centrality bin. A very narrow centrality bin, if considered, would, therefore, minimize this effect but may cause additional \fluc due to statistics. Centrality resolution of the detectors also demands that the chosen centrality bins should not be too narrow. Thus, our observations from FIG.3, tend to suggest that \fluc effects dominate if the centrality bin width is somewhat larger or quite small.\\   
\begin{figure}[htbp]
\begin{center}
\includegraphics[width=12cm,height=8cm]{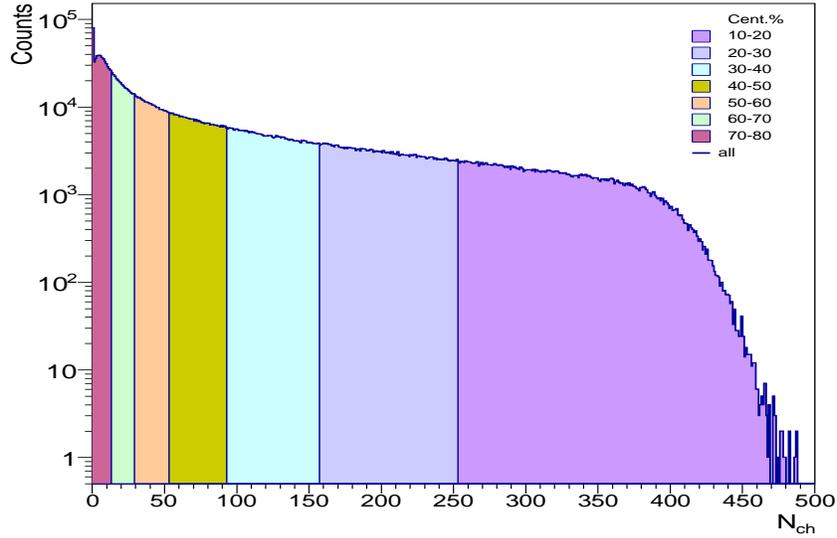}
\end{center}
\caption{An example of centrality selection from the multiplicity distribution of minimum bias simulated events at E${lab}$ = 40A GeV.}
\end{figure}
\begin{figure}[htbp]
\begin{center}
\includegraphics[width=12cm,height=14cm]{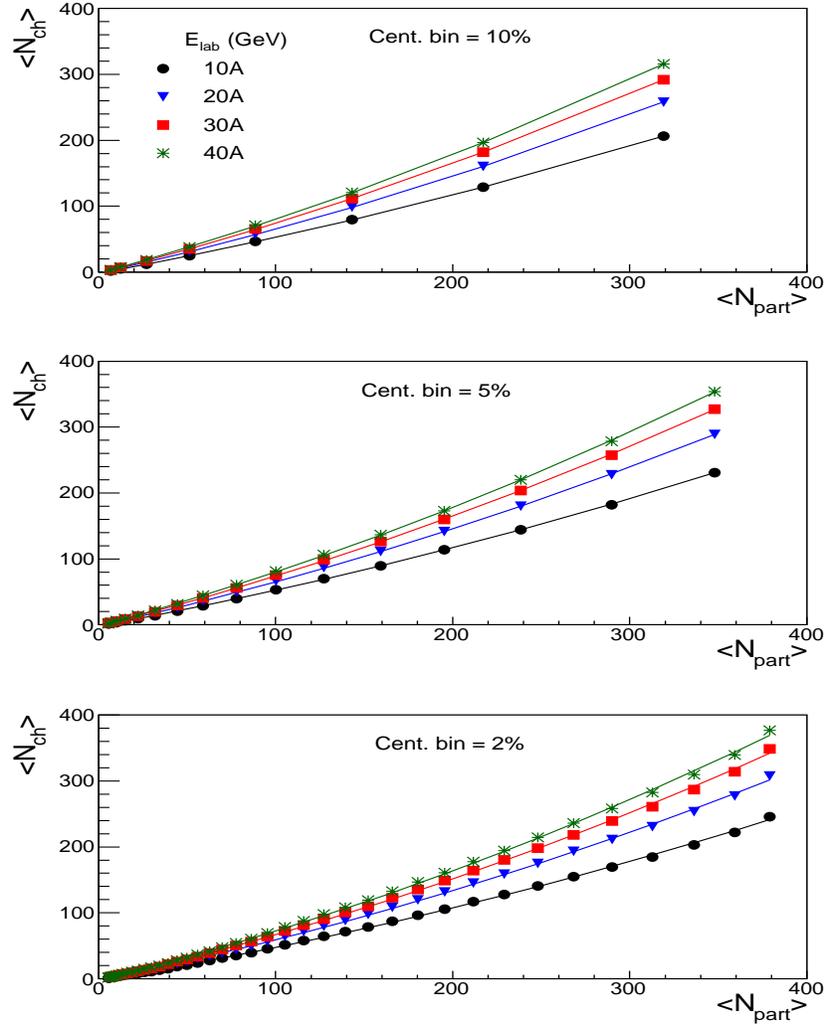}
\end{center}
\caption{Variations of mean multiplicity with mean number of participating nucleons. The lines are due to fits obtained using Eq.1.}
\end{figure}

\begin{figure}[htbp]
\begin{center}
\includegraphics[width=12cm,height=14cm]{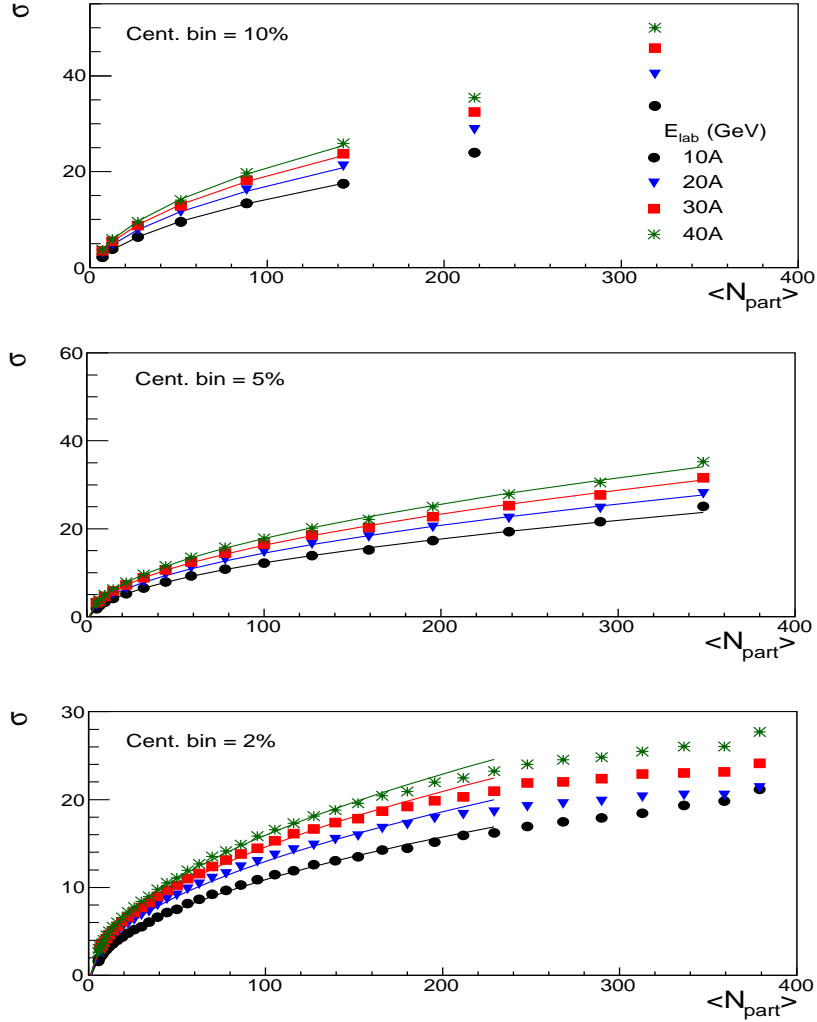}
\end{center}
\caption{Variations of dispersion with mean number of participating nucleons. The lines are due to fits obtained using Eq.2.}
\end{figure}
\begin{figure}[htbp]
\begin{center}
\includegraphics[width=12cm,height=8cm]{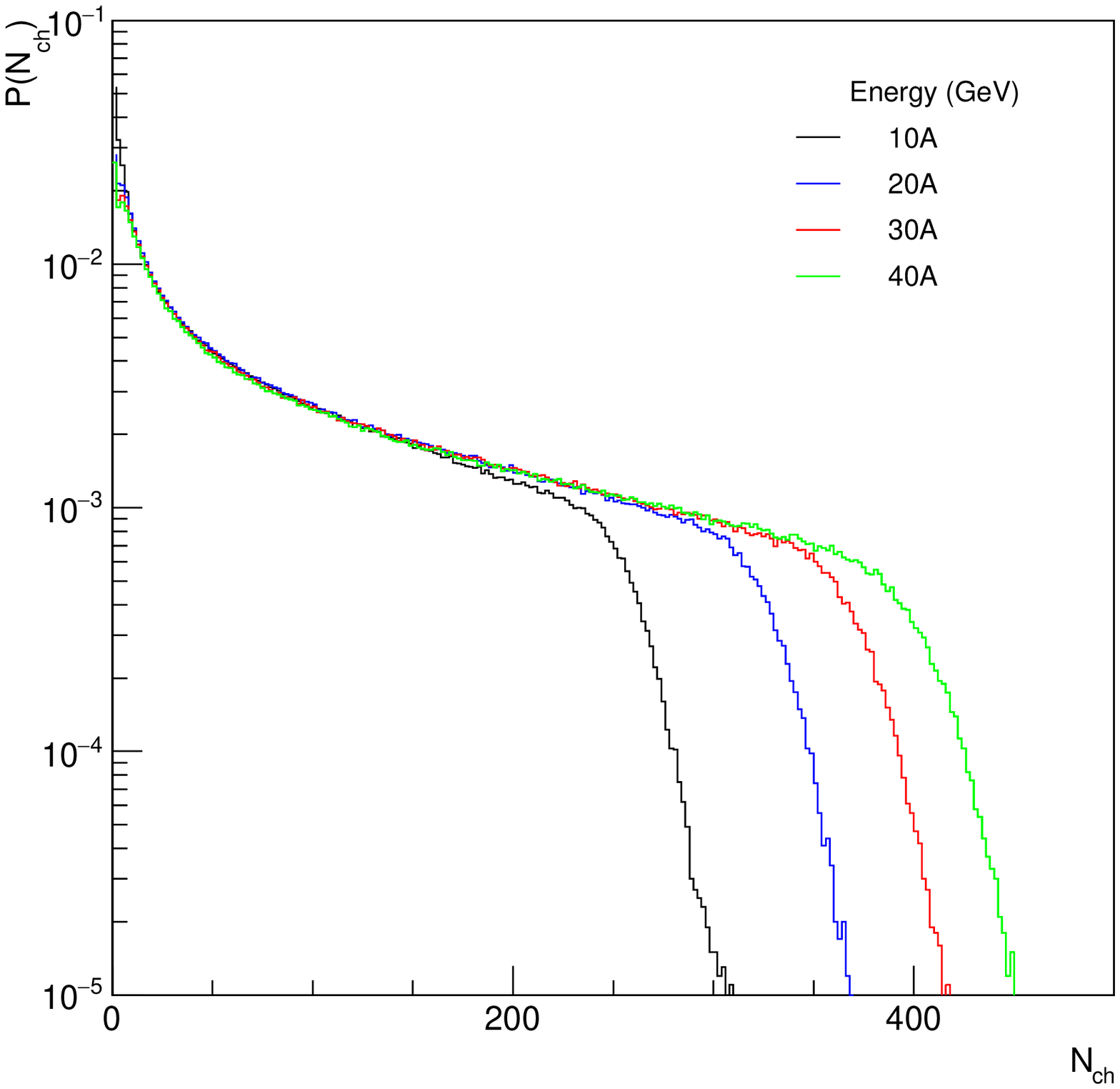}
\end{center}
\caption{Multiplicity distributions of charged particles for AuAu collisions at 10A, 20A 30A and 40A GeV in the range \(p_T = 0.2-5.0\) GeV/c and \(|\eta| = 1.0\).}
\end{figure}
\noindent Multiplicity distributions of relativistic charged particles for minimum bias events for $|\eta| < 1.0$ and $p_{T} = 0.2 - 5.0$ GeV/c  are displayed in FIG.4. It may be noted from the figure that MDs at the four beam energies considered, acquire nearly similar shapes and it is expected that the maximum values of $N_{ch}$ become higher with increasing energies. Similar trends in MDs have also been reported by S. Ghosh et al [17] at the same beam energies predicted  by {\urq} model. MDs of relativistic charged particles for various centrality groups at the four beam energies have also been examined. Distributions for E$_{lab}$ = 40A GeV are presented in FIG.5 along-with the distribution of full sample of events (minimum bias). It is evidently clear from the figure that MD of minimum bias sample is a convolution of MDs with different centrality classes.\\

\begin{figure}[htbp]
\begin{center}
\includegraphics[width=12cm,height=8cm]{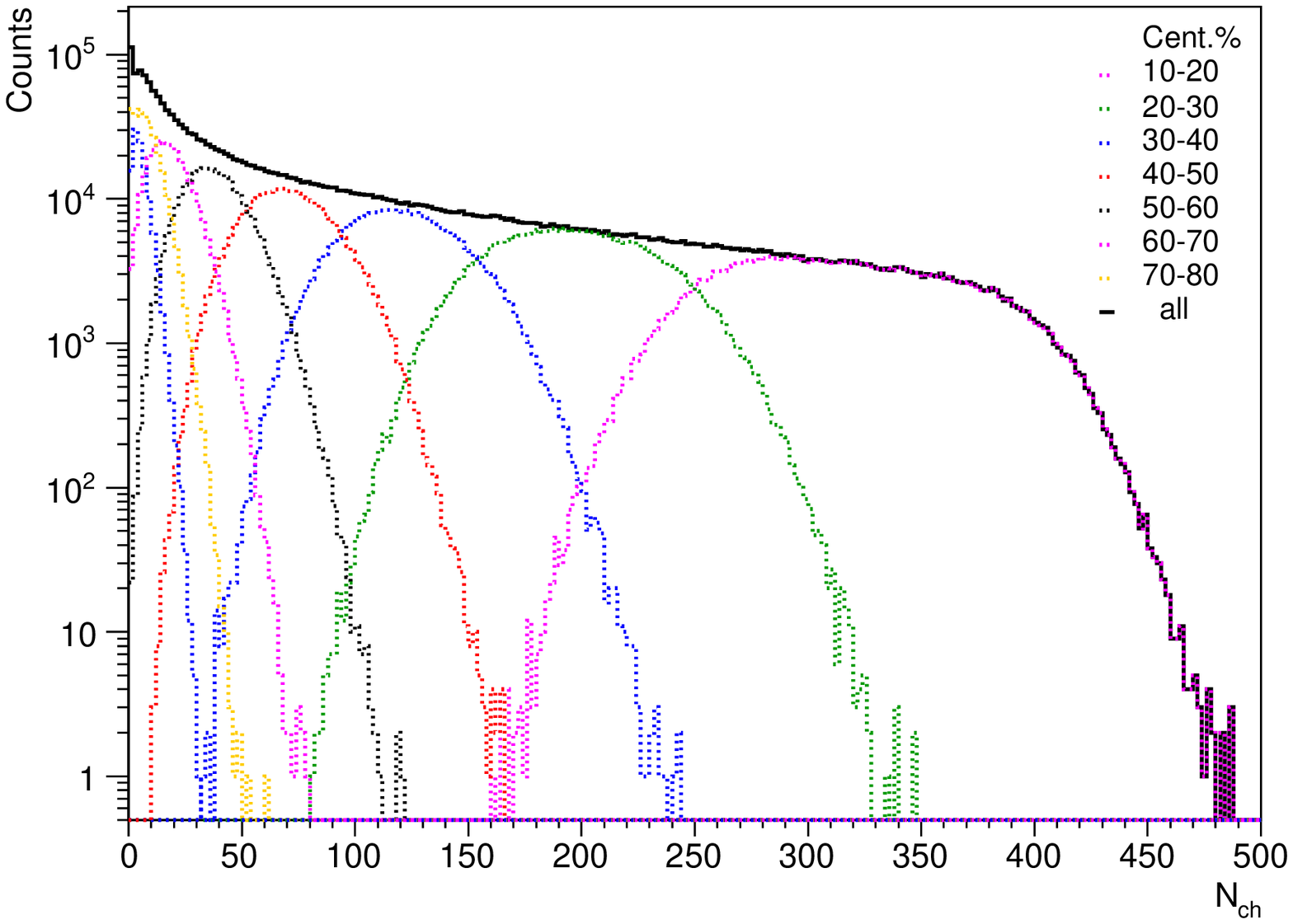}
\end{center}
\caption{Multiplicity distributions of relativistic charged particles for various centrality classes at 40A GeV in the range \(p_T = 0.2-5.0\) GeV/c and \(|\eta| = 1.0\).}
\end{figure}
\begin{figure}[htbp]
\begin{center}
\includegraphics[width=12cm,height=8cm]{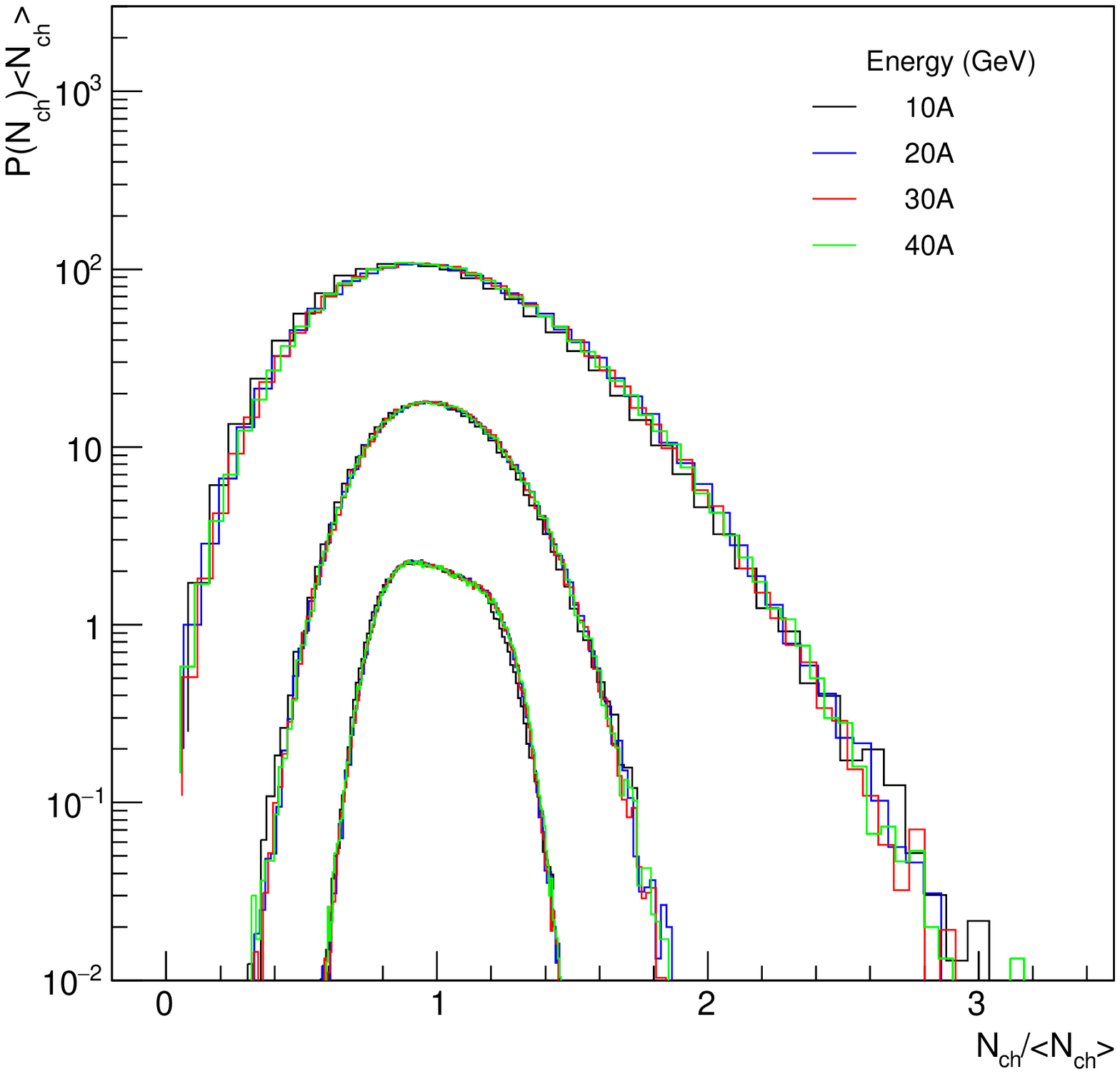}
\end{center}
\caption{Scaled Multiplicity distributions of relativistic charged particles for the centrality bins 0-10\%, 30-40\% and 50-60\%. Distributions corresponding to 30-40\% and 50-60\% are shifted up on the y-scale by factors 10 and 100 for clarity sake only.}
\end{figure}
\noindent Yet another way to examine and predict the MDs, is to plot MDs in terms of KNO scaling variable Z (= $N_{ch}/\langle N_{ch}\rangle$). It has been observed that MDs in hadron-hadron collisions exhibit a universal behavior in a wide range of incident energies if plotted as $\langle N_{ch}\rangle P(N_{ch})$ against the variable Z [20,21,22,23,24,25]. It was shown that MDs corresponding to pp collisions in the energy range $\sim$ (50 - 303) GeV are nicely reproduced by the functional form given by Slattery [22]. MDs in pp collisions, for non single diffractive events at ISR energies have also been observed to exhibit KNO scaling [26]. Since the width of MDs for a given centrality gives the extent of fluctuations, the origin of the \fluc are, thus, inherent in the width of MDs. To understand this behavior, MDs should be plotted for different centrality bins in terms of KNO scaling variable. MDs for 10, 30 and 50$\%$ centrality are plotted in terms of KNO scaling variable in FIG.6. For clarity sake, each next distribution is shifted up on y-scale by a factor of 10. It is observed that the distributions become wider with increasing collision centrality, but exhibits a perfect scaling behavior. MDs, plotted in terms of KNO variable for full event sample in FIG.7, are also noticed to show a perfect KNO scaling. \\
\begin{figure}[htbp]
\begin{center}
\includegraphics[width=12cm,height=8cm]{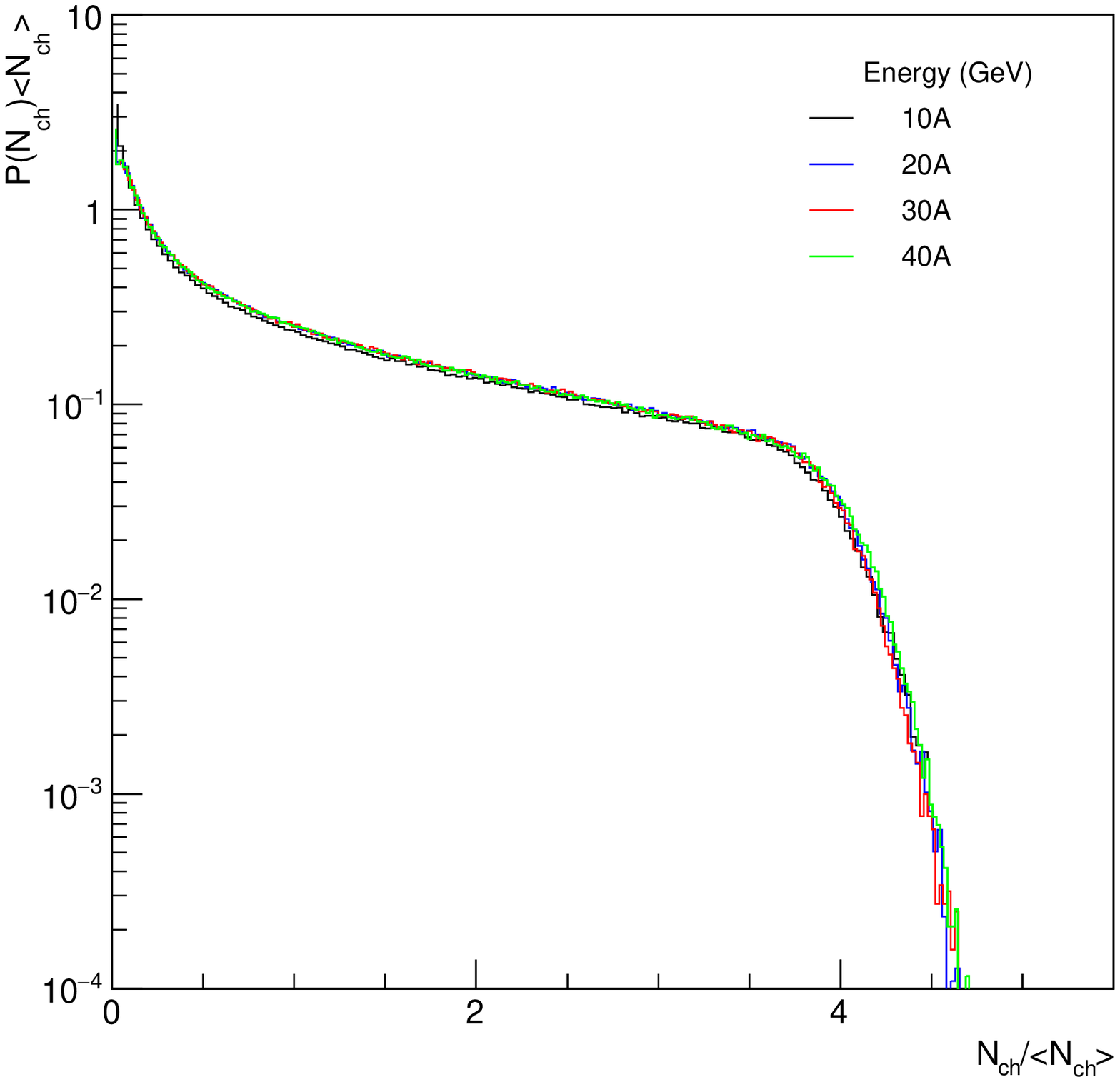}
\end{center}
\caption{Scaled Multiplicity distributions of relativistic charged particles for the minimum bias events at the four beam energies considered.}
\end{figure}
\begin{figure}[htbp]
\begin{center}
\includegraphics[width=12cm,height=8cm]{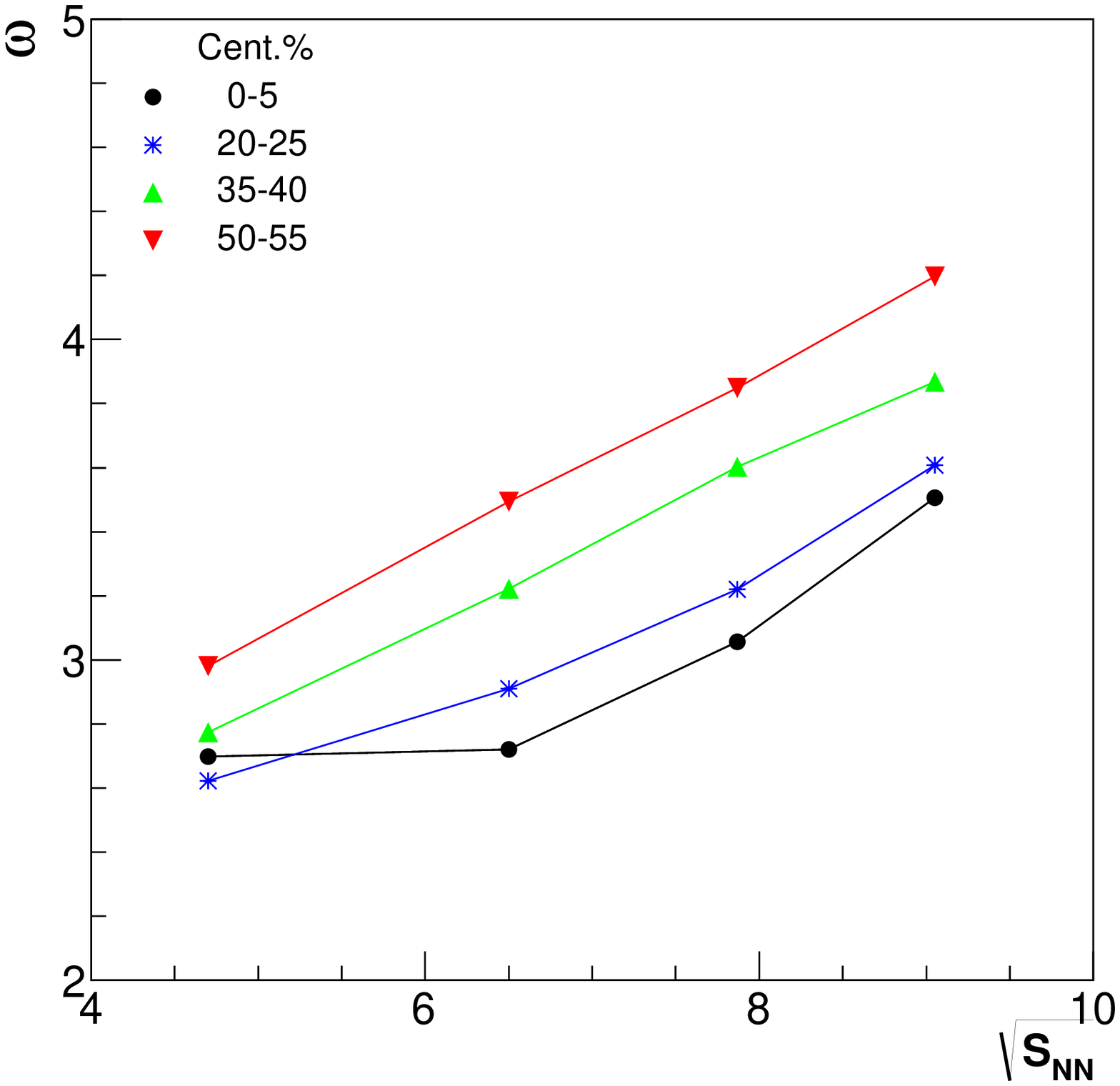}
\end{center}
\caption{Dependence of scaled variance, $\omega$ on beam energy.}
\end{figure}
\noindent The scaled variance, $\omega$ of the MDs defined as,
\begin{equation}
\omega = \frac{\sigma^2}{\langle N_{ch}\rangle}
\end{equation}
here $\omega$ is regarded as a quantitative measure of the particle number fluctuations [7,18,27,28,29]. The scaled variance, $\omega$
 is an intensive quantity which does not depend on the volume of the system within the grand canonical ensemble (GCE) of statistical mechanics or on the number of sources within models of independent source, like wounded nucleon model. The value of scaled variance will be zero in the absence of fluctuations in MDs and unity for  Poisson MDs. Since the volume of the system created in AA collisions fluctuates from event to event, and $\omega$ would depend on volume fluctuations, it becomes important to reduce the fluctuation effects in fluctuation studies [28]. As mentioned earlier, one way to reduce the fluctuation effects is to reduce the number of participating nucleons by selecting the narrow centrality bins. However, the choices of centrality should be such that it does not introduce additional fluctuations due to finite multiplicity and detector resolutions. Once the statistical fluctuation part is under control, the fluctuation effects present will be mostly of dynamical origin which may contain interesting physics associated with the collisions, like hydrodynamic expansion, hadronization at freeze-out, etc.\\

\begin{figure}[htbp]
\begin{center}
\includegraphics[width=12cm,height=14cm]{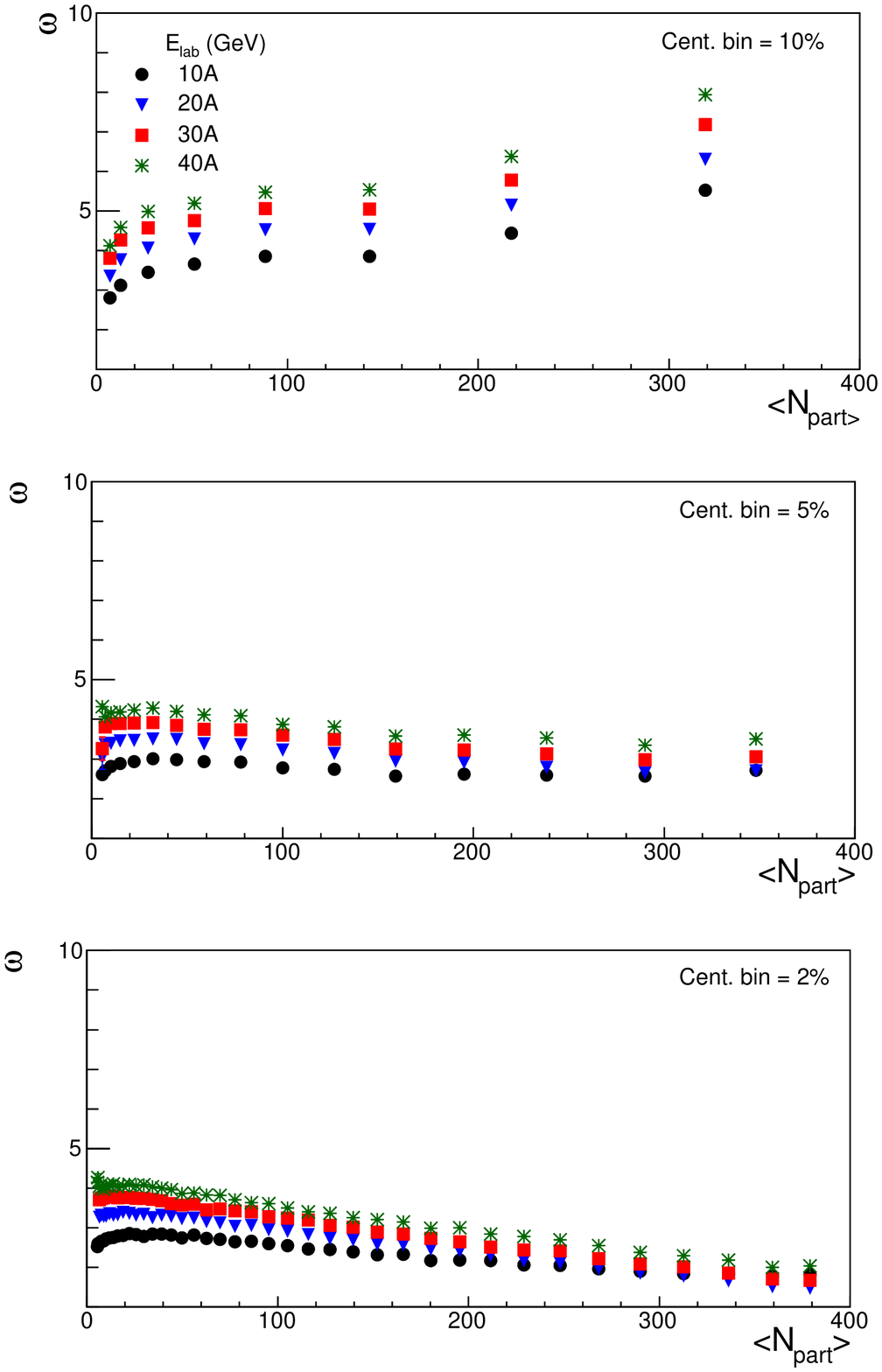}
\end{center}
\caption{Dependence of scaled variance, $\omega$ on \(<N_{part}>\) in different centrality-bin widths.}
\end{figure}

\noindent Variation of scaled variance with c.m. energy  for different centrality bins are plotted in FIG.8. It may be noted from the figure that $\omega$ increases with beam energy as well as in centrality bin widths. It may also be noted that increase of $\omega$ with c.m. energy becomes linear for the centrality classes 35$\%$ and above. If the data obey the KNO scaling [21], it is predicted that $\omega$ should increase linearly with mean charge multiplicity [29]. It may also be noticed in FIG.8 that increase of $\omega$ with beam energy is somewhat weaker for the central collisions. Similar trends of variations of $\omega$ with energy have also been reported in pp collisions by NA61 collaboration [29]. \\

\noindent Centrality dependence of scaled variance at the four incident energies are exhibited in FIG.9. It is observed that for 10$\%$ centrality bins $\omega$ increases with centrality bin widths, whereas for 5$\%$ and 2$\%$ this parameter slowly decreases with increasing centrality and thereafter tends to acquire nearly constant values. This observation, thus, supports that statistical fluctuations arising due to fluctuations in $N_{part}$ becomes visible if the centrality bin width is $\sim$10$\%$ or more and hence considering a bin as wide as 5$\%$, would help arrive at some meaningful conclusions on dynamical fluctuations, if present.\\
\noindent{\bf 4. Conclusions}\\
 MDs and ebe multiplicity fluctuations in AuAu collisions from the beam energy scan in future heavy-ion experiment at the Facility for Antiproton and Ion Research ({\footnotesize{FAIR}}) are examined  in the frame work of Ultra-Relativistic Quantum Molecular Dynamics model, {\footnotesize{URQMD}}. The mean values of MDs are observed to shift towards the higher multiplicity and the width of the distributions are found to become wider from central to peripheral collisions. The MDs are also observed to obey KNO scaling in various centrality windows as well as for full event (minimum bias) samples. Centrality-bin width dependence of the 2$^{nd}$ moments and scaled variance gives the idea of bin width effect and centrality window-width selection, where the statistical \fluc may be treated as under control.

\newpage
\noindent  {{\bf References}}
\begin{enumerate}
\item[1] S.A. Voloshin, V. Koch and H. G. Ritter, ``Event-by-event fluctuations in collective quantities", {\it Phys. Rev.} C60 (1999) 024091.
\item[2] Shakeel Ahmad et al, ``A study of event-by-event fluctuations in relativistic heavy-ion collisions", {\it Int. J. Mod. Phys.} E23 (2014) 1450065.
\item[3] J.L. Albacete et al, ``The initial state of heavy ion collisions", {\it Int. J. Mod. Phys.} A28 (2013) 1340010.
\item[4] M.A. Stephanov et al, ``Event-by-event fluctuations in heavy ion collisions and the QCD critical point", {\it Phys. Rev.} D60 (1999) 114028.
\item[5] M.A. Stephanov et al, ``Signatures of the tri-critical point in QCD", {\it Phys. Rev. Lett.} 81 (1998) 4816.
\item[6] M.A. Stephanov, ``QCD phase diagram and the critical point", {\it Int. J. Mod. Phys.} A20 (2005) 4387.
\item[7] M. Mukherjee et al, ``Fluctuations in charged particle multiplicities in relativistic heavy-ion collisions", {\it J Phys. G: Nucl. Part. Phys.} 43 (2016) 085102, arXiv:1603.02083v3[nucl-ex].
\item[8] J. B$\ddot a$chler et al (NA35 Collaboration), ``Fluctuations of multiplicities in rapidity windows in sulphur-sulphur collisions at 200A GeV", {\it Z. Phys.} C56 (1992) 347.
\item[9] A. Bialas and R. Peschanski, ``Moments of rapidity distributions as a measure of short-range fluctuations in high-energy collisions", {\it Nucl. Phys.} B273 (1986) 703.
\item[10] R.C. Hwa, ``A proposed analysis of multiplicity fluctuations in high-energy heavy-ion collisions", {\it Phys. Lett.} B201 (1988) 165.
\item[11] R.C. Hwa, ``Enhanced multiplicity fluctuation as a possible signature of quark matter", {\it Prog. Part. Nucl. Phys.} 41 (1988) 277.
\item[12] S.A. Bass et al, ``Microscopic models for ultrarelativistic heavy ion collisions", {\it Z. Phys.} C38 (1998) 255; arXiv:9803035v2 [nucl-th].
\item[13] M. Bleicher et al, ``Relativistic hadron-hadron collisions in the ultra-relativistic quantum molecular dynamics model", {\it J. Phys}. G25 (1999) 1859; arXiv:9909407v1 [hep-ph].
\item[14] S.Z. Belen$'$kji and L.D. Landau, ``Hydrodynamic theory of multiple production of particles", {\it Nuovo Cim. Supp.} 3 (1956) 15.
\item[15] T. Regge, ``Introduction to complex orbital momenta", {\it Nuovo Cim.} 14 (1959) 951.
\item[16] H. St\"ocker and W. Greiner, ``High energy heavy ion collisions--probing the equation of state of highly excited hardronic matter", {\it Phys.Rep.} 137 (1986) 277.
\item[17] S. Ghosh et al, ``Net-charge fluctuation in Au+Au collisions at energies available at the Facility for Antiproton and Ion Research using the UrQMD model" {\it Phys. Rev.} C96 (2017) 024912.
\item[18] V.P. Konchakovski et al, ``Multiplicity fluctuations in nucleus-nucleus collisions: Dependence on energy and atomic number", {\it Phys. Rev.} C76 (2008) 024906, arXiv:0712.2044v2 [nucl-th]
\item[19] M.L. Miller et al, ``Glauber modeling in high-energy nuclear collisions", {\it Annu. Rev. Nucl. Post. Sci.} 57 (2007) 205.
\item[20] A. Shakeel et al, ``Scaling of multiplicity distribution of charged shower particles in proton-nucleus interactions at 400 GeV", {\it Phys. Scr.} 29 (1984) 435.
\item[21] Z. Koba, H. B. Neclson and P. Olesen, ``Scaling of multiplicity distributions in high energy hadron collisions", {\it Nucl. Phys.} B40 (1972) 317.
\item[22] P. Slattery, ``Evidence for the onset of semi-inclusive scaling in proton-proton collisions in the 50 -- 300 GeV/c momentum range", {\it Phys. Rev. Lett.} 29 (1972) 1624.
\item[23] J.W. Martin et al, ``Scaling of multiplicity and angular distributions in p-emulsion interactions at 30, 67 and 200 GeV", {\it Nuovo Cimento} A25 (1975) 447.
\item[24] P. Olesen, ``The 200 GeV multiplicity distribution and scaling", {\it Phy. Lett.} B41 (1972) 602.
\item[25] A.J. Buras et al, ``Multiplicity scaling at low energies, a generalized Wroblewski-formula and the leading particle effect", {\it Phys. Lett.} B47 (1973) 251.
\item[26] A. Breakstone et al, ``Charged multiplicity distribution in pp interactions at CERN ISR energies", {\it Phys. Rev.} D30 (1984) 528.
\item[27] M. Mukherjee, ``Multiplicity distributions and fluctuations in proton-proton and heavy-ion collisions", {\it Euro. Phys J. Web. Conf.} 112 (2016) 04004.
\item[28] Andrey Seryakov, NA61/SHINE Coll., ``Rapid change of multiplicity fluctuations in system size dependence at SPS energies", {\it  KnE Energy and Phys., ICPPA 2017, The 3rd Int. Conf. on Part. Phys. and Astrophy.}, (2018) 170; arXiv:1712.03014v1 [hep-ex].
\item[29] Maja. Mackowiak-Pawlowska and A. Wilczek, NA61 Coll., ``Multiplicity fluctuations of identified hadrons in p+p interactions at SPS energies", {\it J. of Phys Conf. Series} 509 (2014) 012044; arXiv:1402.0707v1 [hep-ex].
\end{enumerate}


\end{document}